\documentclass[12pt]{article}
\usepackage{epsf,amsfonts}

\epsfverbosetrue
\def\double{\mathbb}         

\def\cc{{\double C}}

\def\aa{{\cal A}}
\def\ccc{{\cal C}}
\def\dd{{\cal D}}

\def\hh{{\cal H}}

\def\t{{\rm tr}\,}

\def\ddd{{\,\hbox{$\partial\!\!\!/$}}}

\def\ot{\otimes}
\def\op{\oplus}

\def\bb{\begin{eqnarray}}
\def\ee{\end{eqnarray}}
\def\eee{\nonumber\end{eqnarray}}
\def\pp{\pmatrix}
\def\qq{\quad}

\begin{document}

\hsize 17truecm
\vsize 24truecm
\font\twelve=cmbx10 at 13pt
\font\eightrm=cmr8
\baselineskip 18pt

\begin{titlepage}
${}$
\vskip 3truecm

\centerline{\twelve
DOES NONCOMMUTATIVE GEOMETRY ENCOMPASS \\ }  
\centerline{\twelve LATTICE GAUGE THEORY?}

\vskip 2truecm

\begin{center}
{\bf Meinulf G\"OCKELER,}
\footnote{\, also at
Universit\'e de Provence \qq 
meinulf.goeckeler@physik.uni-regensburg.de}\\
Institut f\"ur Theoretische Physik,\\
Universit\"at Regensburg,\\
D--93040 Regensburg\\
\vskip 1truecm
{\bf Thomas SCH\"UCKER, 
\footnote{\, also at
Universit\'e de Provence \qq 
schucker@cpt.univ-mrs.fr }}\\
Centre de Physique Th\'eorique,\\
 CNRS - Luminy, Case 907\\
F--13288 Marseille Cedex 9

\end{center}

\vskip 1truecm
\leftskip=1cm
\rightskip=1cm
\centerline{\bf Abstract} 

\medskip

We are unable to formulate lattice gauge theories in
the framework of Connes' spectral triples.
 
\vskip 2truecm
PACS-92: 11.15 Gauge field theories\\ 
\indent
MSC-91: 81T13 Yang-Mills and other gauge theories 
 
\vskip 1truecm

\noindent march 1998
\vskip 1truecm
\noindent  TPR-98-17\\
\noindent hep-th/yymmxxx

 \end{titlepage}

\section{Introduction}
 
Our fascination for Connes' noncommutative
geometry \cite{connes} has two sources: 1) It is
general enough to treat continuous and discrete
spaces on equal footing. 2) It has enough structure to
include the Yang--Mills action. A natural question
then is whether noncommutative geometry is
compatible with standard formulations of lattice field
theories \cite{latt}. In a more general
frame of noncommutative geometry starting from
differential algebras Dimakis, M\"uller--Hoissen
\& Striker \cite{dmhs} gave an affirmative answer.
Since then Connes \cite{trip} completed the axiomatic
foundation of noncommutative (Riemannian)
geometry in terms of spectral triples. The triple
consists of an associative involution algebra $\aa$, a
faithful representation $\rho$ on a Hilbert space
$\hh$ and a selfadjoint operator $\dd$ on $\hh$, `the
Dirac operator'. In even dimensions one also requires
the existence of a `chirality', a unitary operator
$\chi$ on
$\hh$ of square one, that anticommutes with $\dd$. A
real spectral triple has furthermore a `real structure',
an antiunitary operator $J$ on $\hh$ of square plus or
minus one. These five items are to satisfy axioms.
These axioms generalize properties of the
commutative spectral triple of Riemannian spin
manifolds $M$. There
$\aa=\ccc^\infty(M)$ is the commutative algebra of
functions on spacetime, $\hh$ is the space of square
integrable spinors, $\dd=\ddd$, $\chi=\gamma_5$ and
the real structure is given by charge conjugation.
These axioms are tailored such that there is a
one--to--one correspondence between commutative
spectral triples and Riemannian spin manifolds. The
items of the spectral triple allow to construct a
differential algebra which is no longer chosen by
hand. For the commutative spectral triple of a
spacetime
$M$ this differential algebra is
isomorphic to de Rham's algebra of differential forms. 

\section{Any spectral triple for a lattice?}

We tried to construct a lattice action in terms of
spectral triples starting from a lattice
version of the Dirac operator. We
failed already at the level of the axioms. More
precisely, it is the first order axiom,
\bb [[\dd,\rho(a)],J\rho(\tilde a)J^{-1}]=0,\qq {\rm
for\ all}\ a,\tilde a\in\aa,\ee 
that puts us out of business. On a smooth spacetime
$M$ this equation just says that the Dirac operator
$\ddd$ is a first order differential operator. 

Consider a finite hypercubic lattice of $N^4$
points labeled by discrete 4-vectors $x$ or $y$. We
take for the algebra 
\bb \aa=\bigoplus_1^{N^4}\cc\ \owns f,\ee
represented on
\bb \hh=\bigoplus_1^{N^4}\cc^k
\ \owns
\psi,\ee by
\bb (\rho(f)\psi)(x):=f(x)\psi(x).\ee
For later purposes we include $k$ additional
degrees of freedom that will be spin or colour.
As a matrix the representation is
$\rho(f)(x,y)=f(x)\delta_{xy}\ot 1_k.$ We take the 
real structure $J$ to be such that 
  $J \rho (f) J^{-1} = \rho (\bar{f})$ with 
  $z\mapsto \bar z$ meaning complex conjugation.
  (In the concrete examples discussed below this will 
  be fulfilled.)
  For the Dirac operator, we keep a general matrix 
  $\dd(x,y)$, where the additional indices running from
1 to $k$ are suppressed. Then the double commutator
in the first order axiom becomes,
\bb [[\dd,\rho(f)],\rho(\bar g)](x,y)=
(f(y)-f(x))(\bar g(y)-\bar g(x))\dd(x,y).\ee
The double commutator vanishes if and only if the
Dirac operator is diagonal in $(x,y)$. This excludes
any kind of difference operator on the lattice as for
instance the Dirac--K\"ahler operator,
\bb \dd(x,y)=\sum_{\mu=1}^4{\textstyle\frac{i}{2}}
\eta_\mu(x)\left[\delta_{y,x+\hat\mu}-
\delta_{y,x-\hat\mu}\right].\ee
Here
\bb \eta_\mu(x):=(-1)^{\sum_{\nu=1}^{\mu-1}x_\nu},
\ee
and $\hat\mu$ denotes the lattice unit
  vector in $\mu$-direction .
We define the chirality by the matrix
\bb \chi(x,y):=\epsilon(x)\delta_{xy},\qq
\epsilon(x):=(-1)^{x_1+x_2+x_3+x_4}.\ee
Then, besides the first order axiom, all other axioms
by Connes are satisfied, in particular Poincar\'e
duality. Indeed, if $\{p_x\}_{x\in N^4}$ is the set of
minimal projectors in
$\aa$ defined by $p_x(y)=\delta_{xy}$, then the
intersection form,
\bb \cap_{xy}:=\t
\left[\chi\,\rho(p_x)\,J\rho(p_y)J^{-1}
\right]=\epsilon(x)\delta_{xy},\ee
is non--degenerate.

On the other hand Poincar\'e duality fails for the
naive lattice Dirac operator acting on $k=4$
component spinors $\psi(x)$,
\bb \dd(x,y)= {\textstyle\frac{i}{2}}\sum_{\mu=1}^4
\left[\delta_{y,x+\hat\mu}-
\delta_{y,x-\hat\mu}\right]\gamma^\mu.\ee 
Here the $\gamma^\mu$ are the four Euclidean,
Hermitian Dirac matrices. The chirality is
$\chi=1_{N^4}\ot \gamma_5$ and the real structure
$J$ is charge conjugation on each lattice point. Then
the intersection form vanishes identically because $\t
\gamma_5=0$.

At this point we recall that Connes' geometric
formulation of the standard model of electro--weak
and strong interaction grew out of the two point
space with Dirac operator
\bb \dd=\pp{0&m\cr m&0},\ee
with $m$ being the inverse distance between the two
points. The two point space is a one dimensional
lattice and this $\dd$ is a kind of lattice Dirac
operator. Still the standard model satisfies the first
order condition and does so by adding antiparticles and
strong interactions. We try to copy this trick:
\bb \aa&=&(\cc\op\cc)\op M_k(\cc)\owns (a,b,c),\\
\hh&=&(\cc\op\cc)\ot\cc^k\,\op\,
(\cc\op\cc)\ot\cc^k,\\ \cr 
\rho(a,b,c)&=&\pp{
\pp{a&0\cr 0&b}\ot 1_k&0\cr 
0&1_2\ot\bar c},\\
\dd&=&\pp{
\pp{0&m\cr m&0}\ot 1_k&0\cr 
0&\pp{0&m\cr m&0}\ot 1_k},\\
\chi&=&\pp{
\pp{-1&0\cr 0&1}\ot 1_k&0\cr 
0&\pp{-1&0\cr 0&1}\ot 1_k},\\
J&=&\pp{
0&1_2\ot 1_k\cr 1_2\ot 1_k}\ \circ\ {\rm
complex\ conjugation}.\ee
Now the first order axiom works because the `colour'
$M_k(\cc)$ is vectorlike. However the Poincar\'e
duality fails, the intersection form is degenerate.
Looking at the standard model, this is not
that astonishing. Our model is necessarily closer to the
standard model with right--handed neutrinos for
which Poincar\'e duality fails as well.

\section{Conclusion}

Our conclusion is short and disappointing. We were
looking for an overlap of lattice field theory and 
noncommutative geometry in the hope it would shed
light on the definition of the functional integral
within Connes' geometry. But so far this
overlap seems to be empty in Connes' strict sense.


\begin{thebibliography}{47}

\bibitem{connes}
 A. Connes, {\it Noncommutative Geometry}, Academic 
Press (1994)
\bibitem{latt}
see, e.g., I. Montvay, G. M\"unster, {\it Quantum Fields
on a Lattice,} Cambridge University Press (1994)
\bibitem{dmhs}
A. Dimakis, F. M\"uller--Hoissen \& T. Striker, 
{\it From continuum to lattice theory via deformation
of the differential calculus}, Phys. Lett. B 300 (1993)
141\\
A. Dimakis, F. M\"uller--Hoissen \& T. Striker,
{\it Non--commutative differential calculus and lattice
gauge theory}, J. Phys. A 26 (1993) 1927
\bibitem{trip}
A. Connes,
{\it Noncommutative geometry and reality}, 
J. Math. Phys. 36 (1995) 6194\\
A. Connes, {\it Gravity coupled with matter and the
foundation of noncommutative geometry},
hep-th/9603053, Comm. Math. Phys. 155 (1996) 109

\end{thebibliography}
 \end{document}